\documentclass[article,twocolumn,english,footinbib,tightenlines,nobibnotes,aps,prl,unsortedaddress,showpacs,superscriptaddress]{revtex4-2}
\usepackage{graphicx}
\usepackage[range-units=single,separate-uncertainty]{siunitx}
\usepackage[color links=true,linkcolor=blue,citecolor=blue,urlcolor=blue]{hyperref}

\usepackage[usenames,dvipsnames]{color}
\usepackage{amsmath}
\usepackage{braket}
\usepackage{amsfonts}
\usepackage{tikz}
\usepackage{amssymb}
\usepackage{mathtools}
\usepackage{pict2e}
\usepackage{graphicx,xcolor}
\usepackage{bbding}
\usepackage{bm}
\usepackage{bbm}
\usepackage{setspace} %%
\usepackage{multirow}
\usepackage{dcolumn}
\usepackage{float}
\usepackage{amsmath,mathrsfs}
\usepackage{amsthm}
\usepackage{epsfig}
\usepackage{array}
\usepackage{color}
\usepackage{ulem}
\usepackage[caption=false]{subfig}
\usepackage{ulem}

\usepackage{comment}

\newcolumntype{P}[1]{>{\centering\arraybackslash}p{#1}}

\newcommand\boldvector[1]{%
  \ifcat\noexpand#1\relax % check if the argument calling a command
    \boldsymbol{#1}
  \else
    \mathbf{#1}% Else it is probably a letter
  \fi
}
\def\vec#1{\boldvector{#1}}
\newcommand{\allo}{\{o_i\}}

\def\frac#1#2{{\textstyle{#1 \over #2}}}

\newcommand{\cre}{{\dag}}

\newcommand{\dd}{\mathrm{d}}

\definecolor{applegreen}{rgb}{0.55, 0.71, 0.0}
\definecolor{battleshipgrey}{rgb}{0.52, 0.52, 0.51}

\begin{document}

% ------------------------------------------------------------------------------
% TITLE AND AUTHORS
% ------------------------------------------------------------------------------

\title{Sublattice modulated superconductivity in the Kagome Hubbard model}

\author{Tilman Schwemmer}
\thanks{These authors contributed equally.}
\affiliation{Institute for Theoretical Physics, Julius-Maximilians-Universität Würzburg, 97074 Würzburg, Germany}
\affiliation{Würzburg-Dresden Cluster of Excellence ct.qmat, Univ. of Würzburg, 97074 Würzburg, Germany}

\author{Hendrik Hohmann}
\thanks{These authors contributed equally.}
\affiliation{Institute for Theoretical Physics, Julius-Maximilians-Universität Würzburg, 97074 Würzburg, Germany}
\affiliation{Würzburg-Dresden Cluster of Excellence ct.qmat, Univ. of Würzburg, 97074 Würzburg, Germany}

\author{Matteo D\"urrnagel}
\thanks{These authors contributed equally.}
\affiliation{Institute for Theoretical Physics, Julius-Maximilians-Universität Würzburg, 97074 Würzburg, Germany}
\affiliation{Würzburg-Dresden Cluster of Excellence ct.qmat, Univ. of Würzburg, 97074 Würzburg, Germany}
\affiliation{Institute for Theoretical Physics, ETH Z\"{u}rich, 8093 Z\"{u}rich, Switzerland}

\author{Janik Potten}
\affiliation{Institute for Theoretical Physics, Julius-Maximilians-Universität Würzburg, 97074 Würzburg, Germany}
\affiliation{Würzburg-Dresden Cluster of Excellence ct.qmat, Univ. of Würzburg, 97074 Würzburg, Germany}

\author{Jacob Beyer}
\affiliation{Institute for Theoretical Physics, Julius-Maximilians-Universität Würzburg, 97074 Würzburg, Germany}
\affiliation{School of Physics, University of Melbourne, Parkville,
    VIC 3010, Australia}
\affiliation{Institute for Theoretical Solid State Physics,
    RWTH Aachen University, 52062 Aachen, Germany}
\affiliation{JARA Fundamentals of Future Information Technology,
    52062 Aachen, Germany}

\author{Stephan Rachel}
\affiliation{School of Physics, University of Melbourne, Parkville,
    VIC 3010, Australia}

\author{Yi-Ming Wu}
\affiliation{Stanford Institute for Theoretical Physics, Stanford
  University, Stanford, California 94305, USA}

\author{Srinivas Raghu}
\affiliation{Stanford Institute for Theoretical Physics, Stanford
  University, Stanford, California 94305, USA}

\author{Tobias M\"uller}
\affiliation{Institute for Theoretical Physics, Julius-Maximilians-Universität Würzburg, 97074 Würzburg, Germany}
\affiliation{Würzburg-Dresden Cluster of Excellence ct.qmat, Univ. of Würzburg, 97074 Würzburg, Germany}

\author{Werner Hanke}
\affiliation{Institute for Theoretical Physics, Julius-Maximilians-Universität Würzburg, 97074 Würzburg, Germany}
\affiliation{Würzburg-Dresden Cluster of Excellence ct.qmat, Univ. of Würzburg, 97074 Würzburg, Germany}

\author{Ronny Thomale}
\email{Corresponding author: rthomale@physik.uni-wuerzburg.de}
\affiliation{Institute for Theoretical Physics, Julius-Maximilians-Universität Würzburg, 97074 Würzburg, Germany}
\affiliation{Würzburg-Dresden Cluster of Excellence ct.qmat, Univ. of Würzburg, 97074 Würzburg, Germany}
\affiliation{Department of Physics and Quantum Centers in Diamond and
  Emerging Materials (QuCenDiEM) group, Indian Institute of Technology
Madras, Chennai 600036, India}

\date{\today}

% ------------------------------------------------------------------------------
% ABSTRACT
% ------------------------------------------------------------------------------

\begin{abstract}
We identify a superconducting order featuring spatial pair modulations on the kagome lattice subject to onsite Hubbard $U$ and nearest
neighbor $V$ interactions. Within our functional renormalization group analysis,
this state appears with a concomitant $d$-wave superconducting (SC) instability at
zero lattice momentum, where it distinguishes itself through \textit{intra-unit cell
modulations} of the pairing function thus breaking the discrete space group symmetry.
The relative weight of the sublattice modulated superconductor (SMS) and $d$-wave SC is influenced by the absolute
interaction strength and coupling ratio $V$/$U$. Parametrically
adjacent to this domain at weak coupling, we find an intra-unit cell
modulated vestigial charge density wave and an $s$-wave SC
instability. Our study provides a microscopic setting and thorough description of this novel SMS arising within a translation symmetry broken background.
\end{abstract}

%%-------------------------------------------------------------------------------

\maketitle

% ------------------------------------------------------------------------------
%  INTRODUCTION
% ------------------------------------------------------------------------------

\paragraph{Introduction.}
Superconducting (SC) order beyond spatially uniform pairing is a longstanding
area of condensed matter
research~\cite{doi:10.1146/annurev-conmatphys-031119-050711}. The idea of a pair
density wave (PDW), i.e., spatially varying electron pairing manifesting
either as finite range fluctuations or long range order has attained significant
attention in the context of high-$T_c$ cuprates~\cite{RevModPhys.87.457}. There,
intriguing phenomena such as Fermi arcs or layer decoupling find a rather
intuitive explanation from the viewpoint of a PDW reference state.
Most recently, PDWs have surfaced in experimental signatures for a plethora
of material classes like kagome metals, transition-metal dichalcogenides,
heavy-fermion compound UTe$_2$ and
pnictides~\cite{roton-pdw, Liu2021, Gu2023, Zhao2023, Li2023}, evoking a
universality of the PDW state in correlated electron systems.
As tempting and rich as the principal phenomenology might be, the microscopic
evidence for a PDW state still is rather poor.
Spatially modulated SC pairing has only been rigorously accessed in the
Fulde-Ferrell-Larkin-Ovchinnikov (FFLO)
scenario of subjecting weak coupling spin-singlet SC to a magnetic
field~\cite{PhysRev.135.A550,LO}. Without an external field, there is no
generic weak coupling solution featuring Cooper pairs at finite lattice
momentum.
Therefore, the majority of previous attempts had to resort to effective
mean field treatments either through postulating finite range bare pair
hopping~\cite{PhysRevB.81.020511} or analyzing higher order terms of Ginzburg
Landau expansions~\cite{bergi,danni,PhysRevX.4.031017}. It is then usually a
matter of non-universal parameter choice whether one finds a PDW as
competing order to uniform SC and charge density wave (CDW), or as a, possibly
fluctuating, high-temperature mother state yielding vestigial SC and CDW
order at lower temperature~\cite{doi:10.1146/annurev-conmatphys-031119-050711}.
\begin{figure*}[t]
    \centering
\includegraphics[width=1.9\columnwidth]{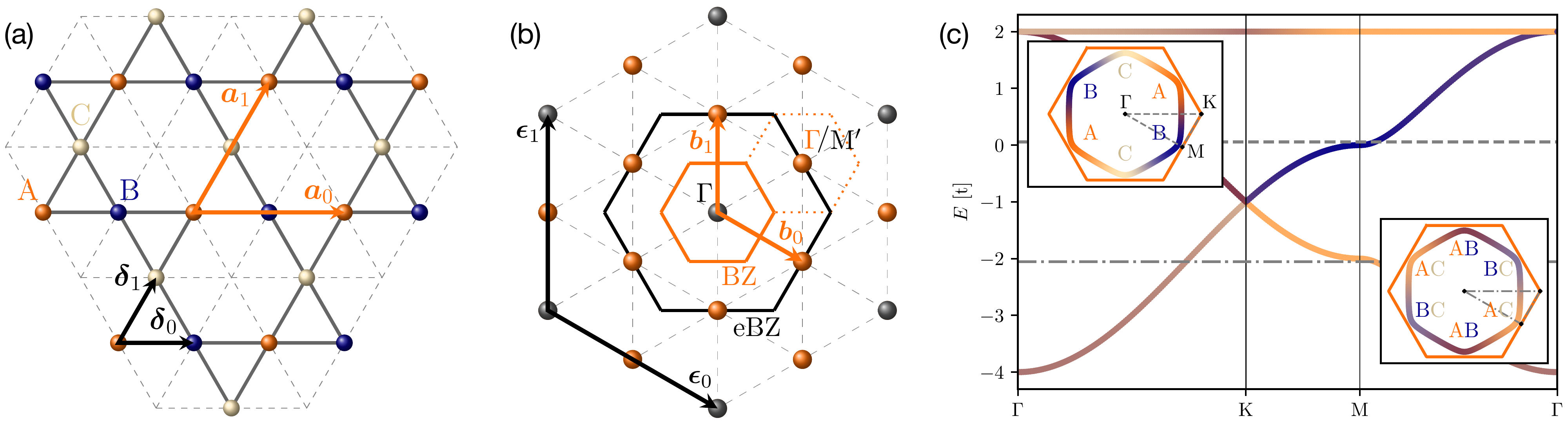}
    \caption{Kagome Fermiology and extended zone scheme. (a) Kagome
      lattice with sublattices $A$, $B$, and $C$ along with the inter
      $\vec a_{0,1}$ and intra $\vec\delta_{0,1}$ unit cell vectors.
      (b) Fourier transform of the kagome lattice with Brillouin zone (BZ) in orange and extended Brillouin zone in black. $M'$ points in the eBZ coincide with the $\Gamma$ point in the BZ.
      (c) The kagome band structure features a flat band and two dispersive bands which display particular eigenstate sublattice distribution according to sublattice interference~\cite{PhysRevB.86.121105}. p-type (m-type) van Hove fillings are symbolized by dashed (dashed-dotted) lines.
      }
    \label{fig:Kagome}
\end{figure*}
It is thus of central interest to identify a microscopic approach to
correlated electron systems that realizes a PDW state~\cite{PhysRevLett.130.026001}.
The functional renormalization group (FRG)~\cite{RevModPhys.84.299, Beyer2022} suggests
itself as a natural method to access ordering instabilities at and beyond weak
coupling. Even though it lacks an analytical control parameter which is the
archetypal challenge for any order at intermediate coupling, the FRG treats all
bilinear fermionic orders on equal footing and allows for a smooth interpolation to
the weak coupling limit by reducing the interaction strength.
FRG calculations were demonstrated to adequately describe even intricate
phase diagrams of multi-orbital / multi-sublattice correlated electron systems
such as iron pnictides~\cite{PhysRevLett.102.047005,doi:10.1080/00018732.2013.862020}.
As a guiding motif for our study of spatially modulated superconductors we identify the non-trivial
sublattice structure of experimentally observed PDW materials.
This suggests the $U$-$V$ kagome Hubbard model (KHM), where $U$ and $V$ denote
the onsite and nearest neighbor (NN) electronic repulsion
on the kagome lattice, respectively, as a blueprint to examine the
sublattice structure of the low energy fluctuations and its imprint on
emergent exotic electronic orders~\cite{PhysRevLett.110.126405,PhysRevB.87.115135}.
In particular, the geometry of the KHM promotes sublattice interference
(SI)~\cite{PhysRevB.86.121105}, leading to pronounced non-local interactions
and emergent NN pair hopping, which creates a
propensity for PDW formation~\cite{si-patch}.

Moreover, the sublattice degree of freedom allows for a formation of intra-unit cell modulated states. 
In analog to an intra-unit cell modulated charge density wave we observe a
superconducting state of equal pair modulation periodicity, that hence features
similar phenomenology to a PDW.
However, this state does not break an additional translation symmetry
but rotational symmetry within the unit cell and hence surfaces as a $\mathbf Q
= 0$ instability in the BZ. As this deviates from the common definition of PDW,
we label the obtained instability sublattice modulated superconductivity (SMS).

In this Letter, we develop a theory for an SMS state observed within FRG calculations.
 Specifically, the state we find does not appear as a parasitic onset order to another
electronic instability, but unfolds as part of the leading instability in the
particle-particle (pp) channel descending from a pristine kagome metallic state.
In terms of symmetry
classification at the pp instability level, admixes with inplane
$d$-wave SC ($d$SC) in the $E_2$ irreducible representation of the hexagonal point group
$C_{6v}$. The key to disentangle the SMS from the $d$SC contribution is the
center of mass dependence of the condensate wave function in the
tri-atomic unit cell basis. Here, the SMS component exhibits intra-unit cell modulations while the $d$SC
component is uniform.

{\it Kagome Hubbard model (KHM).}
We start from
\begin{equation} \label{model}
    H=t\sum_{\langle i,j\rangle} \left
    (c^\dagger_{i\sigma}c^{\vphantom\dagger}_{j\sigma}+\text{h.c.}\right)
    +U\sum_{i=1}^N
    n_{i\uparrow}n_{i\downarrow}+\frac{V}{2}\hspace{-5pt}\sum_{\langle i,j\rangle;\sigma \sigma'} \hspace{-4pt}n_{i\sigma}n_{j\sigma'} \,\text{,}
\end{equation}
where $c_{i \sigma}^\dagger , c_{i \sigma}^{\vphantom\dagger}$ denote electron
creation and annihilation operators at site $i$ with spin $\sigma$,
$n_{i\sigma}=c_{i\sigma}^\dagger c_{i\sigma}^{\vphantom\dagger}$, and $t$, $U$,
and $V$ denote the energy scales of hopping, onsite Hubbard, and NN repulsion, respectively.
The kagome lattice depicted in \autoref{fig:Kagome}a features a non-bipartite lattice of corner-sharing
triangles, resulting in three sublattices $A$, $B$, and $C$ as well as
Bravais lattice vectors $\vec a_0$ and $\vec a_1$.
Alternatively, it can be interpreted as a charge density wave
formation at 3/4 filling of an underlying triangular lattice with halved unitcell
vectors $\vec\delta_{0}$ and $\vec\delta_{1}$, where every fourth unoccupied site is
projected out.
In reciprocal space (\autoref{fig:Kagome}b), this corresponds to the reduced (orange) and extended (black) Brillouin
zone (BZ) spanned by the vectors $\vec b_{0,1}$ and $\vec\epsilon_{0,1}$ according
to $\vec b_i^\intercal \vec a_j =\vec\epsilon_i^\intercal\vec\delta_j =2\pi \delta_{ij}$.
The three inequivalent van Hove points at the M points in the BZ epitomize
a peculiar density of states of differing sublattice occupancy
(pure p-type at filling $n=5/12$ for the upper and
mixed m-type ($n=3/12$) for the lower van Hove level in \autoref{fig:Kagome}c)~\cite{PhysRevB.86.121105,PhysRevLett.127.177001},
a generic feature of kagome metals which has been recently confirmed by angle-resolved photoemission spectroscopy (ARPES)~\cite{kang,hu2021rich}.
For both m-type and p-type, the mismatch of sublattice support for electronic
eigenstates connected by van Hove point nesting triggers SI.

{\it FRG instability analysis.}
The FRG formulates an ultraviolet to infrared cutoff flow of the two-particle
electronic vertex operator $V(\vec k_1,\vec k_2,\vec k_3,\vec k_4) c^\dagger_{\vec k_4} c^\dagger_{\vec k_3}
c^{\vphantom\dagger}_{\vec k_2} c^{\vphantom\dagger}_{\vec k_1}$ as a set of
differential equations, where the $\vec k_i$ denote lattice
momenta.
We employ the FRG in its truncated unity (TU) approximation, where the vertex is separated into
three different contributions according to their distinctive transfer
momenta~\footnote{See Supplemental Material and references \cite{doi:10.1080/00018732.2013.862020,RevModPhys.84.299,LICHTENSTEIN2017100,Schober_2018,PhysRevB.85.035414,Huseman_Honerkamp,metzner_MF1,metzner_MF2,Beyer2022} therein for details on TUFRG and further $E_2$ basis states.},
which can be classified as two particle-hole (ph) and a single particle-particle (pp) channel.
The complementary vertex momenta are expanded in plane wave formfactors $\phi_f(\vec k)
= e^{i \vec r_f \vec k}$ via the Bravais lattice sites $\vec r_f = n
\vec a_0 + m \vec a_1$, $n,m \in \mathbb{Z}$.
For the KHM, any channel vertex function is specified by a 4-tuple of
sublattice indices $\{o_i\}$, $i=1,2,3,4$, transfer momentum $\vec Q$, and two residual momenta which we expand into formfactors according to
\begin{equation} \label{eqn:vertex_ff}
    \mathcal V^{\allo}_{ff'}(\vec Q)
    = \int_{\text{BZ}}\!\!\!\dd \vec k \dd \vec k'
    V^{\allo}(\vec Q,\vec k,\vec k')
    \phi_f(\vec k) \bar\phi_{f'}(\vec k')\,\text{.}
\end{equation}
Truncating the set of basis functions, reflecting the locality of pp
and ph pairs, this unitary expansion becomes approximate.
We terminate the flow at an energy scale $\Lambda_c$, upon encountering a divergence in a vertex element
indicating a symmetry breaking phase transition.
At this point, the effective vertex contains all higher energy quantum fluctuations accessible within the FRG approach.
For SC pairing at $\vec Q=0$, the system's ground state is then obtained from the
BCS gap equation which reduces to an eigenvalue problem
\begin{equation}
    \lambda \Delta_{o_1o_2}^f = \sum_{f' o_3 o_4}
    \mathcal V_{ff'}^{\allo}(0) \, \Delta_{o_3o_4}^{f'}
\label{eq:BCS_orbital}
\end{equation}
at the onset of ordering.
The functional form of the SC order parameter is thus given by the eigenstate $\Delta_{o_1o_2}^f$ associated with the most negative eigenvalue $\lambda$.
Exploiting the group symmetry constraints on
Eq.~\eqref{eq:BCS_orbital} along Schur's lemma allows for a classification of $\Delta_{o_1o_2}^f$ in terms of irreducible representations (irreps) of $C_{6v}$.
Within our TUFRG analysis, we identify an extended region in the
$U$-$V$ parameter space where $\vec Q=0$ instabilities in the spin-singlet pp
channel dominate, and the superconducting order parameter transforms
according to the $E_2$ irrep.

\begin{figure}
    \centering
    \includegraphics[width=\columnwidth]{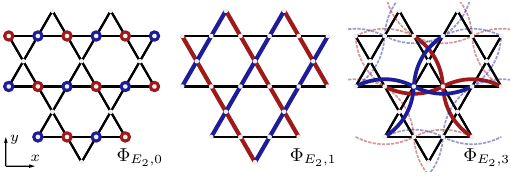}
    \caption{Harmonics of the $E_2$ irrep with
      $y$-mirror eigenvalue $-1$. We show the onsite
      ($\Phi_{E_2,0}$), first NN ($\Phi_{E_2,1}$), and third NN
      harmonic ($\Phi_{E_2,3}$). We skip the second NN due to its
      similarity to the first NN.
      Red/blue indicates positive/negative value of the pair correlation Eq.~\eqref{eq:pair_correlation} between  connected sites.
    }
    \label{fig:measurement}
\end{figure}

{\it Classification of the $E_2$ pp instability.}
A two-electron expectation value in the spin singlet sector
\begin{equation}
    \langle c^\cre_{\vec r_i,\uparrow} c^\cre_{\vec r_j,\downarrow} - c^\cre_{\vec r_j,\uparrow} c^\cre_{\vec r_i,\downarrow} \rangle
    = \delta_{\vec r_i,\vec \rho}\,\delta_{\vec r_j,\vec \rho}\,\Phi(\vec r_i,\vec r_j)
\label{eq:pair_correlation}
\end{equation}
is most conveniently described in terms of a center of mass coordinate (CMC) $\vec R = (\vec r_i + \vec r_j) / 2$ and a relative coordinate (RC)
$\vec r = \vec r_i - \vec r_j$.
In the following, we denote the set of all lattice points by $\vec\rho$ and use $\vec \rho_o$ for points related to sublattice $o \in \{A,B,C\}$. We
introduce $\vec a_2=\vec a_1 - \vec a_0$ and
$\vec\delta_{2}=\vec\delta_{1} - \vec\delta_{0}$ for notational
convenience. Since the kagome lattice is not of Bravais type, the directions
of NN bonds depend on the sublattice.
For each pairing distance, i.e., each value of the RC (onsite, NN, etc.),
the $E_2$ irrep is expanded in pairs of degenerate orthogonal states,
characterized by even or odd transformation behavior under mirror
symmetry along the $y$-axis~\cite{Note1}.
We order different $E_2$ candidate states by their RC length and
constrain ourselves to the mirror-odd state as we present a graphical representation in \autoref{fig:measurement}.
The onsite ($\vec r=0$) pairing wave function reads
\begin{equation}
    \Phi_{E_2,0}(\vec R,\vec r) =
    \Big(
        \delta_{\vec R,\vec \rho_A}
        -\delta_{\vec R,\vec \rho_B}
    \Big)\delta_{\vec r,\vec 0} \,\text{.}
\label{eq:pi_E2_0}
\end{equation}
The absence of a lattice site at the point group's rotation center, allows
for such onsite pairing wave functions on any non-Bravais lattice.
The trivial transformation property of the RC is complemented by a non-trivial CMC function
due to the reduced site symmetry group of the $3c$ Wyckoff position.
In contrast, the NN pairing state
\begin{equation}
    \Phi_{E_2,1}(\vec R,\vec r) =
        \delta_{\vec r,\pm \vec \delta_1}
        -
        \delta_{\vec r, \pm \vec\delta_2}
\label{eq:pi_E2_1}
\end{equation}
features a trivial CMC dependence, and instead it is the RC that captures the $E_2$ transformation behaviour.
Whereas the second NN state is structurally equivalent~\cite{Note1} to the NN state,
the third NN pairing reads
\begin{equation}
    \begin{split}
    \Phi_{E_2,3}(\vec R,\vec r) = \, &
        \delta_{\vec R,\vec \rho_B}
        \delta_{\vec r,\pm\vec a_0}
         -
        \delta_{\vec R,\vec \rho_A}
        \delta_{\vec r,\pm\vec a_0}
        \\
+ \, & \delta_{\vec R,\vec \rho_C}
        \Big(
            \delta_{\vec r,\pm\vec a_1} -
            \delta_{\vec r,\pm\vec a_2}
        \Big).
    \end{split}
\label{eq:pi_E2_3}
\end{equation}
It cannot be decomposed into a product of solely CMC and RC dependent
functions.
This stems from two symmetry-inequivalent types of third NN on the kagome
lattice due to the reduced $C_{2v}$ site symmetry group, that yields the additional
spatial CMC structure~\cite{Note1}.

\begin{figure}
    \centering
    \includegraphics[width=0.48\textwidth]{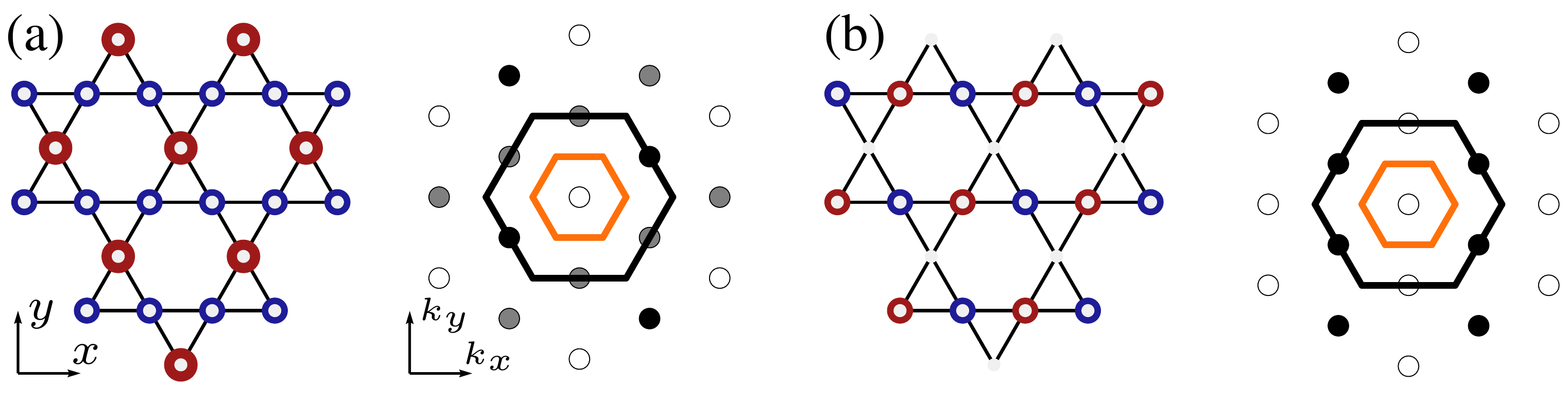}
    \caption{Real space representation of the two onsite pairing wavefunction
    with $E_2$ symmetry and their Cooper pair momentum structure in
    the eBZ. Lighter shading corresponds to less
    intensity. Contrarily to the $d$SC pairing wavefunctions of
    Eq.~\eqref{eq:pi_E2_1}, finite Cooper pair momentum is revealed by the
    unfolding into the eBZ~\cite{Note1}. The depicted $\mathbf Q$ structure
    resembles the expected STM signature of the SMS state.}
    \label{fig:PDW_eBZ}
\end{figure}

{\it SMS wave function.}
Two candidate states may belong to the same irrep and yet trace back to
entirely different physical origins. For the $A_1$ pairing irrep on the square
lattice, this has been extensively discussed in the context of
$s_\pm$ vs. $s_{++}$-wave SC in iron pnictides~\cite{Hirschfeld_2011,doi:10.1146/annurev-conmatphys-020911-125055}.
For the kagome lattice, the fundamentally different nature of $E_2$ formfactors becomes apparent upon disentangling the spatial information encoded in CMC and RC, respectively.
The Cooper pair wave function of typical zero momentum SCs features a uniform CMC dependence, while the RC encodes the transformation behavior of the associated irrep.
Eq.~\eqref{eq:pi_E2_0}, however, describes a scenario where the
typical roles of CMC and RC are interchanged.
It obeys $\sum_{\vec R} \Phi(\vec R,\vec r) = 0$, which is a characteristic of PDW-type
pairing~\cite{Arovas2022}.
The key point to appreciate here is that despite the $\vec Q=0$
channel,
{\it spatially modulated contributions can be distinguished from homogeneous ones within the multi-site unit cell}.
This is the essential difference between the SMS and $d$SC contributions to an $E_2$ pp instability in the KHM.

Restoring the full real space information previously encoded in the
sublattice indices of the tri-atomic basis by transfering the SMS state of
Eq.~\eqref{eq:pi_E2_0} to the extended BZ shown in~\autoref{fig:PDW_eBZ} reveals finite momentum Cooper
pairing,
that can be directly attributed to a generalised version of the structure
factor experimentally accessible in STM measurements (for details cf.
SM~\cite{Note1}):
By locally probing the differential conductance via the
Josephson tunneling signal of a superconducting STM tip, the SC gap can be
monitored with high spatial resolution. The Fourier transform of the pairing gap
maps exhibits peaks at the wave vectors associated with the Cooper pair modulation,
that can be matched with the peak structure in~\autoref{fig:PDW_eBZ}. This allows for a clear experimental distinction between the proposed SMS state
and uniform SC pairing~\cite{PhysRevB.64.212506, Pan1998, Proslier2006, PhysRevB.80.144506, Karan_2022}.

\begin{figure}
    \centering
    \includegraphics[width=0.48\textwidth]{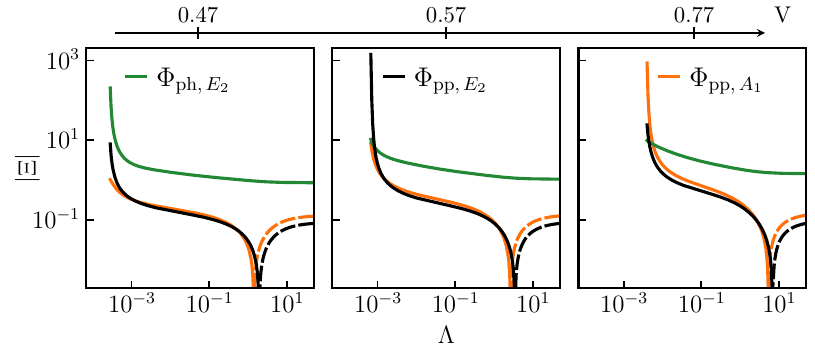}
    \caption{FRG vertex analysis for
      $U/t=0.14$ and \mbox{$0.47 < V/t < 0.77$} as a function of the energy cutoff $\Lambda$.
      $E_2$~CDW ($\Phi_{\text{ph},\, E_2}$, green), SMS $E_2$ pp order
      ($\Phi_{\text{pp},\, E_2}$, black), and SC $A_1$ pp
      ($\Phi_{\text{pp}, \,A_1}$, orange) compete for the leading instability.
      Attractive (repulsive) expectation values $|\Xi|$ of the leading instabilities at $\Lambda_c$ are indicated by solid (dashed) lines~\cite{Note1}.
    }
    \label{fig:motherstate}
\end{figure}

{\it SMS phase domain.}
Within our TUFRG analysis of the KHM, the $E_2$ pp instability domain is governed by intra-sublattice pairing, i.e., the eigenstates of Eq.~\eqref{eq:BCS_orbital} are described by a linear combination of $\Phi_{E_2,0}$ and $\Phi_{E_2,3}$~\cite{Note1}.
This allows to identify the $E_2$ pp instability as a SMS.
Such states were previously found, yet not adequately appreciated in the
kagome phase diagram of Ref.~\onlinecite{PhysRevB.87.115135, PhysRevB.105.075118}.
The intra-sublattice pairing tendency can be understood from the
viewpoint of SI:
Under the RG, effects of $U$ are suppressed while $V$
grows~\cite{si-patch}, making inter-sublattice pairing less favourable.
This trend also manifests in the fact that the SMS instability at small $U/t$ is parametrically framed by an $E_2$-type CDW and an $A_1$-type, i.e., $s$-wave SC (\autoref{fig:motherstate}) at p-type van Hove filling.

The SMS state emerges as interstitial order via competing interaction processes projected into the pairing channel:
$U$ drives non-local spin fluctuations;
$V$, by contrast, generates charge fluctuations entering the Cooper channel as onsite attractive interactions \cite{Romer_2022}.
In \autoref{fig:motherstate}, we illustrate the generic scenario of
competing instabilities generated by this interplay of local and
non-local interactions by comparing FRG flows for varying bare values
$0.47< V < 0.77$ at fixed $U=0.14$, i.e., located in the
weak coupling domain.
Both SMS-type and $s$-wave SC channels are initially repulsive (dashed lines) and subsequently turn attractive (solid lines) through the RG flow.
Remarkably, the adjacent CDW features the same irrep and intra-sublattice modulation as the SMS phase, suggesting vestigial symmetry-type restoration at the phase transition.
In previous contexts, PDW fluctuations were similarly found to predominantly occur around phase transitions between a density wave phase and SC pairing, where a variety of competing and/or cooperative orders arise at intermediate coupling strength~\cite{doi:10.1146/annurev-conmatphys-031119-050711}.
Any weak coupling theory of PDW has hitherto relied on fine-tuned Fermiology~\cite{wc_PDW}.
Here, however, we find that the SMS prevails beyond p-type van Hove filling whenever intra-unit cell modulations become competitive to homogeneous pairing functions due to SI.

\begin{figure}
    \centering
    \includegraphics[width=0.48\textwidth]{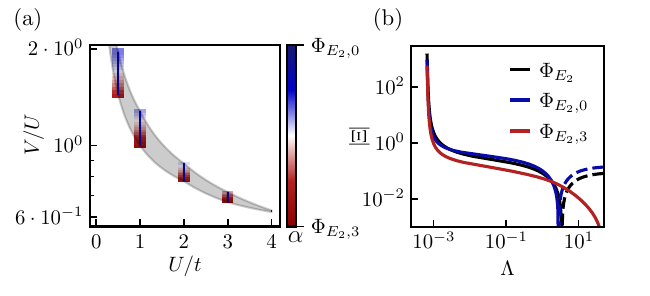}
    \caption{
    (a)  Admixture ratio $\alpha$ of $\Phi_{E_2,0}$ (blue)~\eqref{eq:pi_E2_0} and $\Phi_{E_2,3}$ (red)~\eqref{eq:pi_E2_3}
    for the $E_2$ pp domain at p-type van Hove filling ($n=5/12$). (b) Expectation value of  the
    leading eigenstate at $\Lambda_c$ (black), $\Phi_{E_2,0}$, and
    $\Phi_{E_2,3}$. Negative (positive) values are indicated by solid (dashed) lines.
     }
    \label{fig:phasespace}
\end{figure}

The competition of interaction processes is directly reflected in the
pairing wave functions overlap with $\Phi_{E_2,0}$ and
$\Phi_{E_2,3}$~\cite{Note1} and their relative weight $\alpha$, which
is depicted in \autoref{fig:phasespace}a for the SMS regime at p-type van Hove filling. Doping slightly away from this point significantly increases the SMS region, e.g. by a factor of $\approx 2.5$ for $U=3$ at $n=1.01 \times 5/12$.
The third NN state $\Phi_{E_2,3}$ does not suffer an energy penalty from $U$ in~\eqref{model} and thus promotes attractive pairing channel already at bare coupling (\autoref{fig:phasespace}b).
Analogous to the previous discussion, $V$-driven intra-unit cell charge fluctuations
yield an onsite attraction of Cooper pairs through the RG flow~\cite{PhysRevB.87.115135}.
The onsite contribution $\Phi_{E_2,0}$ to SMS increases with $V$, and eventually drives the system into an $s$-wave SC.

% ------------------------------------------------------------------------------
%  CONCLUSION
% ------------------------------------------------------------------------------
\paragraph{Conclusion and outlook.}
Our findings emphasize the fundamental role of the
sublattice texture on the Fermi surface in a microscopic theory of spatially
modulated pairing, a motif also shared among the different PDW candidates~\cite{wc_PDW, Zhou2022, Jiang2023}.
The new type of sublattice modulated SC instability we find for the KHM model, is facilitated
by the nontrivial basis in the unit cell. This allows spatial modulations of the
Cooper pair wave function, while still exploiting the Cooper logarithm of a $\mathbf Q = 0$
pairing instability as main driver for the symmetry breaking transition.
This establishes the proposed type of SMS as a generic instability of a Fermi
liquid with sublattice degrees of freedom and may explain the broad range of
materials exhibiting PDW phenomenology.
The breaking of rotational symmetry instead of translational symmetry by
intra-unit cell modulations of the SC order parameter finds a direct
correspondence for magnetically ordered systems in the new field of
altermagnetism~\cite{PhysRevX.12.040002,PhysRevX.12.040501}.
Viewing the kagome lattice as a ficticious 3/4 filling CDW of an underlying
triangular lattice (\autoref{fig:Kagome}a), we recover the expected intertwined
character of PDW: the kagome-CDW at $2 \vec\delta_{1,0}$ comes along with a $2 \vec\delta_{1,0}$
PDW competing with $d$SC.
This motif carries over to SC within the $2\times2$ charge order in 135 kagome compounds~\cite{roton-pdw,Zhou2022}: Spatially modulated intra-unit cell SC can arise within the 
 enlarged $2\times2$ unit cell in the charge ordered phase without an additional translation symmetry breaking imposed by the SC 
 transition. Within the SMS phase, experimental signatures of PDW like reciprocal Josephson effect can still be observed~\cite{Daido2022, He2022, PhysRevLett.130.126001}.
To target experimentally observed PDW states in kagome compounds like the
roton density wave reported for CsV$_3$Sb$_5$ within the
charge order domain~\cite{roton-pdw}, the methodological shortcoming to surmount
in the future is to continue the FRG flow throughout the preceding charge ordered
phases~\cite{Salmhofer2004}.

%-------------------------------------------------------------------------------

% ------------------------------------------------------------------------------
%  ACKNOWLEDGEMENT
% ------------------------------------------------------------------------------

\paragraph{Acknowledgement.}
We thank S. A. Kivelson, M. Klett, J.B. Hauck, and M. Sigrist for discussions.
The work is funded by the Deutsche Forschungsgemeinschaft (DFG, German Research Foundation) through Project-ID 258499086 - SFB 1170 and and through the W\"urzburg-Dresden Cluster of Excellence on Complexity and Topology in Quantum Matter -- \textit{ct.qmat} Project-ID 390858490 - EXC 2147.
We acknowledge HPC resources provided by the Erlangen National High Performance Computing Center (NHR@FAU) of the Friedrich-Alexander-Universit\"at Erlangen-N\"urnberg (FAU). NHR funding is provided by federal and Bavarian state authorities. NHR@FAU hardware is partially funded by the DFG – 440719683.
M.D. received funding from the European Research Council under Grant No. 771503 (TopMech-Mat).
S.R. was supported in part by the US Department of Energy, Office of Basic Energy Sciences, Division of Materials Sciences and Engineering, under contract number DE-AC02-76SF00515.
J.B. thanks the DFG for support through RTG 1995 from SPP 244 ``2DMP''.

%\paragraph{Author contributions.}
%-------------------------------------------------------------------------------

% \bibliography{pdw}
%
\end{document}

% ------------------------------------------------------------------------------
% TITLE AND AUTHORS
% ------------------------------------------------------------------------------

\title{Supplemental Material for \\ Sublattice modulated
superconductivity in the Kagome Hubbard Model}

  \author{Tilman Schwemmer}
  \thanks{These authors contributed equally.}
  \affiliation{Institute for Theoretical Physics, Julius-Maximilians-Universität Würzburg, 97074 Würzburg, Germany}
  \affiliation{Würzburg-Dresden Cluster of Excellence ct.qmat, Univ. of Würzburg, 97074 Würzburg, Germany}

  \author{Hendrik Hohmann}
  \thanks{These authors contributed equally.}
  \affiliation{Institute for Theoretical Physics, Julius-Maximilians-Universität Würzburg, 97074 Würzburg, Germany}
  \affiliation{Würzburg-Dresden Cluster of Excellence ct.qmat, Univ. of Würzburg, 97074 Würzburg, Germany}

  \author{Matteo D\"urrnagel}
  \thanks{These authors contributed equally.}
  \affiliation{Institute for Theoretical Physics, Julius-Maximilians-Universität Würzburg, 97074 Würzburg, Germany}
  \affiliation{Würzburg-Dresden Cluster of Excellence ct.qmat, Univ. of Würzburg, 97074 Würzburg, Germany}
  \affiliation{Institute for Theoretical Physics, ETH Z\"{u}rich, 8093 Z\"{u}rich, Switzerland}

  \author{Janik Potten}
  \affiliation{Institute for Theoretical Physics, Julius-Maximilians-Universität Würzburg, 97074 Würzburg, Germany}
  \affiliation{Würzburg-Dresden Cluster of Excellence ct.qmat, Univ. of Würzburg, 97074 Würzburg, Germany}

  \author{Jacob Beyer}
  \affiliation{Institute for Theoretical Physics, Julius-Maximilians-Universität Würzburg, 97074 Würzburg, Germany}
  \affiliation{School of Physics, University of Melbourne, Parkville,
      VIC 3010, Australia}
  \affiliation{Institute for Theoretical Solid State Physics,
      RWTH Aachen University, 52062 Aachen, Germany}
  \affiliation{JARA Fundamentals of Future Information Technology,
      52062 Aachen, Germany}

  \author{Stephan Rachel}
  \affiliation{School of Physics, University of Melbourne, Parkville,
      VIC 3010, Australia}

  \author{Yi-Ming Wu}
  \affiliation{Stanford Institute for Theoretical Physics, Stanford
    University, Stanford, California 94305, USA}

  \author{Srinivas Raghu}
  \affiliation{Stanford Institute for Theoretical Physics, Stanford
    University, Stanford, California 94305, USA}

  \author{Tobias M\"uller}
  \affiliation{Institute for Theoretical Physics, Julius-Maximilians-Universität Würzburg, 97074 Würzburg, Germany}
  \affiliation{Würzburg-Dresden Cluster of Excellence ct.qmat, Univ. of Würzburg, 97074 Würzburg, Germany}

  \author{Werner Hanke}
  \affiliation{Institute for Theoretical Physics, Julius-Maximilians-Universität Würzburg, 97074 Würzburg, Germany}
  \affiliation{Würzburg-Dresden Cluster of Excellence ct.qmat, Univ. of Würzburg, 97074 Würzburg, Germany}

  \author{Ronny Thomale}
  \email{Corresponding author: rthomale@physik.uni-wuerzburg.de}
  \affiliation{Institute for Theoretical Physics, Julius-Maximilians-Universität Würzburg, 97074 Würzburg, Germany}
  \affiliation{Würzburg-Dresden Cluster of Excellence ct.qmat, Univ. of Würzburg, 97074 Würzburg, Germany}
  \affiliation{Department of Physics and Quantum Centers in Diamond and
    Emerging Materials (QuCenDiEM) group, Indian Institute of Technology
  Madras, Chennai 600036, India}

\date{\today}

% ------------------------------------------------------------------------------
% ABSTRACT
% ------------------------------------------------------------------------------

%%-------------------------------------------------------------------------------

\maketitle

\section{Truncated Unity FRG}
While the natural energy scale of the non-interacting theory is determined by the bandwidth of the
single particle spectrum, the situation becomes more delicate in the presence of interactions.
Interaction effects establish collective ordering phenomena at energy scale several orders of magnitude
below the strength of the bare interaction in the initial Hamiltonian due to screening by the
non interacting background.
To trace down the effective contribution of interactions to the physics at the important energy scales,
a variety of renormalization procedures have been developed to incorporate higher order screening processes
in the low energy effective theory.

Among these, the functional renormalization group stands out by its unbiased differentiation of phases
and is the natural method for the analysis of spatially modulated states of matter like pair density waves (PDW) or sublattice modulated superconducting (SMS) orders as proposed in the manuscript. While we omit the basics of FRG itself by referencing to
the literature \cite{doi:10.1080/00018732.2013.862020,RevModPhys.84.299} instead, we want to reintroduce
the TUFRG \cite{LICHTENSTEIN2017100,Schober_2018} approximation here:
We start out with the integro-differential equation of the functional
renormalization group in its SU(2) reduced form:
\begin{figure}[b]
    \includegraphics[width=\columnwidth]{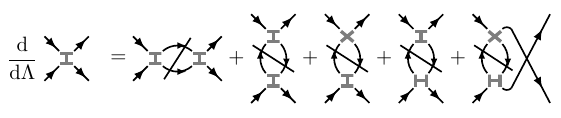}
    \caption{Diagrammatic representation of the SU$(2)$-TUFRG
    flow.
    The dash through the pair of propagator lines indicates the
    derivative w.r.t. the scale $\Lambda$.
    The diagrams can be separated into the three channels by their
    transfer momentum between the 2-particle vertices:
    The first diagram is associated with the $P$-channel, the next three
    contribute to the $D$-channel and $C$-channel consists of the last diagram.}
    \label{diag:flow_diag}
\end{figure}
The exact perturbative expansion of the FRG vertex has already been truncated at the two-particle
level to make the infinite hierarchy of coupled equations solvable.

We furthermore exploit spin rotational invariance, which allows to consider the opposite spin channel only.
The symmetric and antisymmetric part of the thereby obtained effective interaction resembles the singlet
and triplet correlations respectively.
For brevity we drop the spin and sublattice indices in all following formulae, which we retain as tensor
dimensions of the object $V^{o_1o_2o_3o_4}_{\uparrow\downarrow\uparrow\downarrow}(\vec k_1, \vec k_2, \vec k_3)$.
The exact order of indices can be inferred from the diagrammatic expression where the shape
of the vertices indicates connected sublattices.

We further employ the truncated unity expansion to the
FRG which is based on the singular mode paradigms~\cite{PhysRevB.85.035414}.
To this end the vertex is decomposed in the
\textit{particle-particle (P)}, \textit{direct particle-hole (D)},
\textit{crossed particle-hole (C)} channel
\begin{equation}
    V = V_P + V_D + V_C \,
    \label{eqn:V_split}
\end{equation}
according to the three possible diagrammatic transfer momenta
$\vec Q_P = \vec k_1 + \vec k_2$,
$\vec Q_D = \vec k_1 - \vec k_3$ and
$\vec Q_C = \vec k_1 - \vec k_4$.
This primary momentum dependence drives any divergence in the channel,
so the remaining
spatial dependence of any given function $F(\vec k_1, \vec k_2, \vec k_3)$ can be expanded in a
suitable
set of formfactors $\phi_{f}(\vec k)$ via
\begin{multline}
        \hat{X}[F]_{ff'}(\vec Q_X) =
        \int \dd \vec k_X \int \dd \vec k'_X \phi_{f}(\vec k_X)\phi_{f'}(\vec k'_X) \\
        F\left(\vec k_1(\vec Q_X, \vec k_X, \vec k'_X),
        \vec k_2(\vec Q_X, \vec k_X, \vec k'_X),
        \vec k_3(\vec Q_X, \vec k_X, \vec k'_X) \right)\, , \\
\label{eq:ff_expansion}
\end{multline}
where $X\in\{P,C,D\}$.
Because of the slow-varying nature of the interaction w.r.t. the secondary
momenta, the high frequency harmonics can be truncated without significant loss
of information.
In the present study we choose plane waves $\phi_{f}(\vec k) = e^{-i \vec k \vec r_f}$ as a complete
basis with $\vec r_f$ representing a Bravais lattice site and consider sites up to third nearest
neighbors.
By applying the above projection to \autoref{diag:flow_diag}, the RG equation for the vertex
\begin{equation}
    \frac {d}{d\Lambda}\hat X[\mathcal V] =\hat X[\dot{\mathcal V}_P] + \hat X[\dot{\mathcal V}_D] + \hat X[\dot{\mathcal V}_C] \,
    \label{eqn:dV}
\end{equation}
breaks down to the three separate channel flow equations
\begin{align}
\begin{split}
    \hat P [\dot{\mathcal V}_P]_{ff'} (\vec Q_P)
    &=  \sum_{f_1,f_2}
            \mathcal V_{ff_1}(\vec Q_P)
            \dot L^{pp}_{f_1f_2}(\vec Q_P)
            \mathcal V_{f_2f'}(\vec Q_P) \\
    \hat D [\dot{\mathcal V}_D]_{ff'} (\vec Q_D)
    &=  \sum_{f_1,f_2}
        \left[ -2
        \mathcal V_{f f_1}(\vec Q_D)
        \dot L^{ph}_{f_1 f_2}(\vec Q_D)
        \mathcal V_{f_2 f'}(\vec Q_D) \right. \\
    &\qquad
       +
        \mathcal V_{f f_1}(\vec Q_C)
        \dot L^{ph}_{f_1f_2}(\vec Q_D)
        \mathcal V_{f_2f'}(\vec Q_D) \\
    &\qquad \left.
        +
        \mathcal V_{ff_1}(\vec Q_D)
        \dot L^{ph}_{f_1f_2}(\vec Q_D)
        \mathcal V_{f_2f'}(\vec Q_C)
        \right]\\
    \hat C [\dot{\mathcal V}_C]_{ff'} (\vec Q_C)
    &=  \sum_{f_1,f_2}
            \mathcal V_{ff_1}(\vec Q_C)
            \dot L^{ph}_{f_1f_2}(\vec Q_C)
            \mathcal V_{f_2,f'}(\vec Q_C) \,.
        \label{eqn:flow}
\end{split}
\end{align}
Here $\dot L^{pp}_{ff'}(\vec Q)$ and $\dot L^{ph}_{ff'}(\vec Q)$ are the
product of single-scale propagator and propagator
$(\frac{d G^{\Lambda}G^{\Lambda}}{d\Lambda})$ in either particle-particle (pp) or
particle-hole (ph) configuration transformed into TU representation.

Since renormalisation effects for each transfer momentum are  accumulated
separately, the computational complexity of the calculations reduces
significantly while retaining momentum conservation which was not possible in
the n-patch FRG schemes used previously \cite{doi:10.1080/00018732.2013.862020}.
Additionally, the full orbital and band space dependencies are kept throughout the
RG procedure, since the expansion in Eq.~\eqref{eq:ff_expansion} only effects the momentum quantum numbers.
In particular for systems with sublattice, this allows to extract the real space structure within the unit cell
from the order parameter at the end of the flow.

We are left with the remaining problem of the cross-channel projections,
\textit{i.e.} the contribution of each $\dot V_X$ in the two complementary
channels.
By respecting the interplay between the channels, the TUFRG distinguishes itself
from the calculation of three simultaneous but disconnected RPA flows and becomes an
unbiased method, which treats propensities towards particle-particle
instabilities on equal footing to particle-hole instabilities.
The cross channel projections can be reduced to complicated tensor
transformations for each pair $X,Y$ from $\{P,C,D\} : X \neq Y$
\begin{multline}
        \dd \mathcal V^{\allo}_{X,ff'}(\vec Q_X) \\
    = \sum_{\vec Q_Y} M^{X,ff'}_{Y,f_1f_2}(\vec Q_X, \vec Q_Y)
        \dd \mathcal V^{\mathcal P_{X,Y}[\allo]}_{Y,f_1f_2} (\vec Q_Y) \, .
\end{multline}
This set of integro-differential equations and coupling equations
can be integrated as it would be in ordinary FRG:

To obtain a log scale theory close to the Fermi level, we start from the free
theory at $\Lambda \sim \infty$ and integrate the flow equations towards lower
scale.
Upon encountering a phase transition one (or more) of the
differentials in Eq.~\eqref{eqn:flow} diverges at energy scale $\Lambda_C$, which
can be related to the critical temperature $T_C \propto \Lambda_C$ of the
transition.
To associate the divergent channels with physical instabilities,
the effective interaction at $\Lambda_C$ is transformed back into momentum space and
decomposed into the pairing
\begin{equation}
\label{eq:PP_vertex}
\begin{split}
        \Gamma_\text{pp} = & \sum_{\vec Q \vec k \vec k'} \sum_{o_i}
        V^{\allo}_P\big(\vec Q, \vec k, \vec k' \big) \\
      & c^\cre_{\vec k + \vec Q/2,o_2 \uparrow}
        c^\cre_{-\vec k + \vec Q/2,o_3 \downarrow}
        c^\ann_{-\vec k' + \vec Q/2,o_1 \downarrow}
        c^\ann_{\vec k' + \vec Q/2,o_0 \uparrow},
\end{split}
\end{equation}
charge
\begin{equation}
\label{eq:Charge_vertex}
\begin{split}
        \Gamma_{\text{cha}} = & \sum_{\vec Q \vec k \vec k'} \sum_{o_i s s'}
        \left[ V^{\allo}_{D}\big(\vec Q, \vec k, \vec k' \big) -
               \frac{1}{2}
               V^{o_0o_1o_3o_2}_C\big(\vec Q, \vec k, \vec k' \big)
        \right]\\
      & c^\cre_{\vec k - \vec Q/2,o_2 s}
        c^\ann_{\vec k + \vec Q/2,o_0 s}
        c^\cre_{\vec k' + \vec Q/2,o_3 s'}
        c^\ann_{\vec k' - \vec Q/2,o_1 s'},
\end{split}
\end{equation}
and magnetic
\begin{equation}
\label{eq:Spin_vertex}
\begin{split}
        \Gamma_\text{mag} = & \sum_{\vec Q \vec k \vec k'} \sum_{\{o_i\} \{s_i\}}
        - V^{\allo}_{C}\big(\vec Q, \vec k, \vec k' \big)
        \sigma^z_{s_0s_2} \sigma^z_{s_1s_3}\\
      & c^\cre_{\vec k' + \vec Q/2,o_2 s_2}
        c^\ann_{\vec k + \vec Q/2,o_0 s_0}
        c^\cre_{\vec k -  \vec Q/2,o_3 s_3}
        c^\ann_{\vec k' - \vec Q/2,o_1 s_1}
\end{split}
\end{equation}
contributions \cite{LICHTENSTEIN2017100,Huseman_Honerkamp}.
Each renormalized interaction in momentum space is obtained from the quantities in Eq.~\eqref{eqn:flow}
via the inverse of Eq.~\eqref{eq:ff_expansion}, $\sigma^z$ is the third Pauli matrix.
Since the magnetic channel is not further discussed in the weak coupling domain of SMS elaborated on in the main article, ``cha'' is abbreviated by ``ph'' for notational convenience.

\subsection{Mean field analysis}

From the effective vertex in the most divergent channel, the functional dependence of the order
parameter can be analysed by a subsequent mean field (MF) decomposition of the residual
interaction term \cite{doi:10.1080/00018732.2013.862020,metzner_MF1,metzner_MF2}.
Subsequently, we focus on the pairing channel at $\vec Q_P = 0$, which characterizes uniform superconductivity
and sublattice modulated superconductivity. An analogous procedure yields the order parameter of adjacent phases
like charge and spin density waves and bond orders.

We start by introducing the non vanishing expectation value
\begin{equation}
    \Delta(\vec k) = \frac{1}{N} \sum_{\vec k'} \Gamma_P(\vec k, \vec k') \langle c^\ann_\downarrow(-\vec k') c^\ann_\uparrow(\vec k') \rangle
\end{equation}
with $\vec k, \vec k'$ restricted to energy scales below the cutoff, \textit{i.e.} the single particle energies satisfy $|\xi_n(\vec k)| < \Lambda_C \sim 10^{-4} t$.
Assuming no accidental band degeneracies at the Fermi level, this allows a unique mapping between momentum and band quantum number and we hence drop the latter one.
Neglecting second order fluctuations around the MF solution, the free energy of the system can be obtained in terms of the eigenspectrum of the Bogoliubov quasiparticles
$\xi(\vec k) = \sqrt{E(\vec k)^2 + |\Delta(\vec k)|^2}$ as
\begin{equation}
\begin{split}
    F = & - T \sum_{\vec k} \ln \left( 1 + e^{- \beta \xi(\vec k)} \right)
        - \frac{1}{2} \sum_{\vec k} \left( \xi(\vec k) - E(\vec k) \right) \\
        & + \frac{1}{2} \sum_{\vec k} \frac{|\Delta(\vec k)|^2}{2 \xi(\vec k)}
          \tanh \left( \frac{\xi(\vec k)}{2 T} \right) \, .
\end{split}
\label{eq:free_energy}
\end{equation}
Its minimization w.r.t. the order parameter gives the BCS gap equation
\begin{equation}
    \Delta(\vec k) = - \frac{1}{N} \sum_{\vec k'} \Gamma_P(\vec k, \vec k')
     \frac{\Delta(\vec k')}{2 \xi(\vec k')}
     \tanh \left( \frac{\xi(\vec k')}{2 T} \right) \, .
\label{eq:BCS_gap}
\end{equation}
Since the FRG flow breaks down at the onset of a phase transition, the appropriate temperature scale at the end of the flow is given by $T_C$, which implies $\Delta \rightarrow 0$.
Thus, the righthand side of the above equation can be expanded to first order in the order parameter. From the linearised version of Eq.~\eqref{eq:BCS_gap},
\begin{equation}
    \lambda \Delta(\vec k) = \frac{1}{N} \sum_{\vec k'} \Gamma_P(\vec k, \vec k') \Delta(\vec k') \, ,
\label{eq:lin_BCS_gap}
\end{equation}
the spatial dependence of the order parameter is given by the eigenstate of
$\Gamma_P$ associated with the most negative eigenvalue $\lambda$.
In the case of degenerate leading pairing eigenvalues, the system will realise the complex superposition of the degenerate modes, that minimizes the free energy in Eq.~\eqref{eq:free_energy}.

\subsection{Vertex symmetrization routine}
\label{sec:SymRoutine}

\begin{figure}
    \includegraphics[width=\columnwidth]{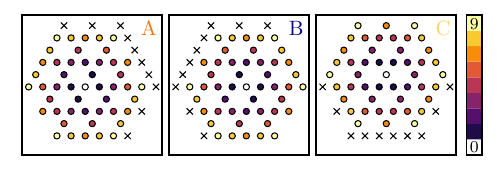}
    \caption{Resymmetrization of formfactor expansion.
    When calculating in proper gauge we implicitly mix different distances in the truncation of the formfactor expansion in Eq.~\ref{eq:ff_expansion}.
    To circumvent the thereby induced symmetry breaking, the formfactor basis has to be adjusted for each sublattice individually to suffice the site symmetry group $C_{2v}$.
    Sites {\sffamily x} do not have a symmetry partner w.r.t the central site and are therefore discarded.
    }
    \label{fig:symmetrization}
\end{figure}

For computational convenience, the FRG treatment of the effective vertex is usually performed in proper gauge as describe in Ref.~\cite{Beyer2022}:
To maintain a periodic Hamiltonian w.r.t. the reciprocal lattice vectors, we choose a gauge where all sites $o$ within the unit cell ($o\in
\{$A,B,C$\}$) are transformed respective the origin of the unit cell
\begin{equation}
    c_{i, o} = \int_{\text{BZ}}\! d\vec{k} e^{\mathrm{i}\vec{k}\vec{r}_i}c_{\vec{k}, o}\,,
\end{equation}
with $\vec{r}_i=n\vec{a}_0+m\vec{a}_1,$ and $n,m\in\mathbb{Z}$
as opposed to the improper and non-periodic form
\begin{equation}
    c_{i, o} = \int_{\text{BZ}}\! d\vec{k} e^{\mathrm{i}\vec{k(}\vec{r}_i+\vec{r}_o)}c_{\vec{k}, o}\,,
\end{equation}
where they are transformed respective their position. $\vec{r}_o$ denotes the shift from the unit cell origin to the position of sublattice $o$. Thus, $(\vec{r}_i + \vec{r}_o )$ are all lattice vectors of the Kagome lattice.
The formfactor expansion in \autoref{eq:ff_expansion} considers only a finite number of formfactors, which is defined by the number of formfactor shells ($n_\text{shell}$) around the onsite unit cell. This induces a truncation of neighboring real-space sites $\{\vec{r}_i: |\vec{r}_i| \leq d_n\}$, where $d_n$ is the distance between onsite- and $n_\text{shell}$-NN unit cell. This is sufficient for intra-orbital terms.
However, for inter-orbital contributions the different sublattice positions $\vec{r}_o$ within a unit cell mix differing length scales \cite{Beyer2022}.
Now $\exists\, \vec{r}_{i, o} : |\vec{r}_i + \vec{r}_o | > d_n$ and those elements break the symmetry of the system as the unit cell orientation favors one direction. An illustration of all sites contained in the formfactor expansion up to third-NN unit cells can be found in \autoref{fig:symmetrization}.
To restore the symmetry we set elements that do not have a symmetry partner w.r.t. the central site (white) to zero in both, the scale derivatives of the propagators, and the effective interactions in each iteration of the FRG flow.

\section{Complementing basis states of the $E_2$ irrep}
\begin{figure}
    \centering
    \includegraphics[width=\columnwidth]{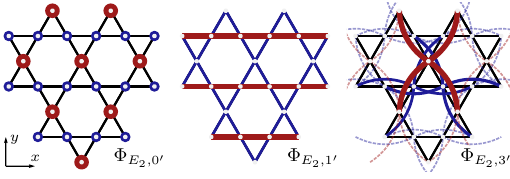}
    \caption{
    Orthogonal complement of the onsite ($\Phi_{E_2,0'}$), nearest neighbor ($\Phi_{E_2,1'}$), and third nearest neighbor ($\Phi_{E_2,3'}$) pairing states presented in \autoref{fig:measurement} with $y$-mirror eigenvalue $+1$ (``even’’).
    Red/blue indicates positive/negative value of the pair correlation between connected sites.
    The thickness of the lines symbolizes the magnitude of pair correlations (red: +2, blue: -1).
    For $\Phi_{E_2,3'}$ only the connections to the central unit cell are highlighted to preserve visual clarity.
     }
    \label{fig:evenStates}
\end{figure}
In the main text the set of all lattice sites is referred to as $\vec \rho$ while subsets are denoted as $\vec \rho_i$. With this we can define the lattice constraint factors $\delta_{\vec R, \vec \rho}$ as
\begin{align}
\delta_{\vec R,\vec \rho}\equiv\sum_{n,m\in\mathbb{Z}}\sum_o\delta_{\vec R, n\vec a_0 + m\vec a_1 + \vec r_o},
\end{align}
where $\vec r_o$ with $o\in A,B,C$ denotes the positions of the atoms $A,B$ and $C$ in the unit cell with $n,m\in\mathbb{Z}$. The subset of one type of sublattice thus follows as
\begin{align}
\delta_{\vec R, \vec \rho_o}\equiv\sum_{n,m\in\mathbb{Z}}\delta_{\vec R, n\vec a_0 + m\vec a_1 + \vec r_o}.
\end{align}
Another relevant subset consists of the corresponding closest centerpoints of a hexagon to a lattice site. We will denote it as $\vec\rho_A\pm\vec\delta_2$, $\vec\rho_B\pm\vec\delta_1$ and $\vec\rho_C\pm\vec\delta_0$, which are always the sites not present in the kagome lattice. For the $A$ sites this results in
\begin{align}
\delta_{\vec R, \vec \rho_A\pm\vec\delta_2}\equiv\sum_{n,m\in\mathbb{Z}}\delta_{\vec R, n\vec a_0 + m\vec a_1 + \vec r_A\pm\vec{\delta_2}}.
\end{align}
The other two terms follow similarly.
Various other subsets can be constructed in a similar fashion if needed.
With this we are able to separate the inherent translation breaking of the kagome lattice via $\delta_{\vec{r}_k,\vec{\rho}}$. The additional spatial structure induced by the emergent phase unveils the character of the instability.
 \begin{figure*}[t]
    \centering
    \includegraphics[width=2.0\columnwidth]{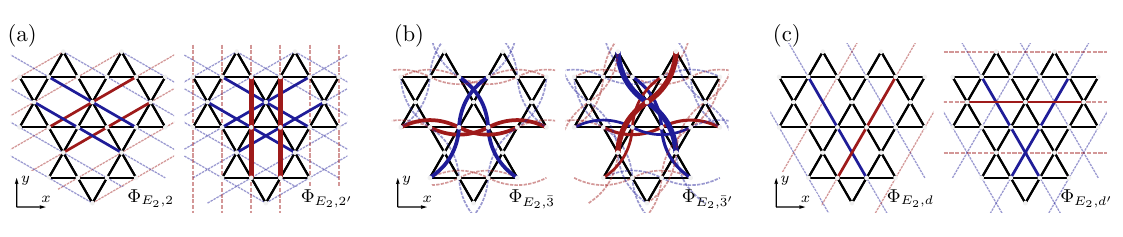}
    \caption{
    Complementing $E_2$ pairing states of the kagome lattice up to third nearest-neighbor. (a) Next nearest-neighbor paring states with odd ($\Phi_{E_2,2}$) and even ($\Phi_{E_2,2'}$) transformation property w.r.t. a $y$-mirror.
    (b) and (c) correspond to third nearest-neighbor paring states whereas ($\Phi_{E_2,\Bar3}$) and ($\Phi_{E_2,\Bar3'}$) are formed by connections along the chains of the lattice while ($\Phi_{E_2,d}$) and ($\Phi_{E_2,d'}$) establish a link across the hexagons.
    Red/blue indicates the sign (+/-) for the pair correlation. The magnitude
    of pair correlations ($\pm 1$ or  $\pm 2$) is symbolized by the
    thickness of the lines.
   Connections to the central unit cell are highlighted for visual clarity.
    }
    \label{fig:E22States}
\end{figure*}
In the main text we chose the mirror symmetry along the $y$-axis as a tool to differentiate the two degenerate states transforming under the $E_2$ irrep. In detail, this mirror symmetry maps all $C$ sites onto themselves while $A$ and $B$ sites are switched. From this, one can determine two additional symmetry operations that could equally have been chosen for our classification, namely $A\leftrightarrow C$ or $B\leftrightarrow C$. Since they are symmetry equivalent it suffices to only consider one of them.

\subsection{Harmonic expansion of the $E_2$ sector}

In this section we show the orthogonal states corresponding to the ones displayed in the main paper and complement these with the remaining states transforming under the $E_2$ irrep up to third nearest neighboring sites.
A detailed analysis of their contributions is given which validates our restriction to $\Phi_{E_0}$ and $\Phi_{E_3}$ as the main contributions of the FRG state.
%Since we only covered mirror odd $\Phi_{E_2}$ states with a substantial contribution to the PDW phase in the FRG analysis, we will discuss the missing basis states within the first three nearest neighbor shells in the following.
The next nearest neighbor correlation state
\begin{equation}
    \Phi_{E_2,2}(\vec R,\vec r) =
        \delta_{\vec r, \pm(\vec\delta_0+\vec\delta_1)}
        -
        \delta_{\vec r,\pm(\vec\delta_0-\vec\delta_2)},
\label{eq:pi_E2_2}
\end{equation}
is structurally similar to $\Phi_{E_2,1}$,
since it only consists of a relative coordinate (RC) dependence due to the center of mass coordinate (CMC) being already accounted for by the lattice constraint.
In the third nearest neighbor shell, the number of sites exceeds the rank of the point group,
which promotes two additional types of third nearest neighbor states:
\begin{equation}
	\begin{split}
		\Phi_{E_2,\Bar3}(\vec R,\vec r) = \, &
		\delta_{\vec R,\vec \rho_B}
		\delta_{\vec r,\pm\vec a_0}
		+
		\delta_{\vec R,\vec \rho_A}
		\delta_{\vec r,\pm\vec a_0}
		\\
		- \, & \delta_{\vec R,\vec \rho_C}
		\Big(
		\delta_{\vec r,\pm\vec a_1} +
		\delta_{\vec r,\pm\vec a_2}
		\Big).
	\end{split}
	\label{eq:pi_E2_3Bar}
\end{equation}
is symmetric under a mirror operation along the $y$-axis and thus perpendicular to $\Phi_{E_2,3}$.
The second state
\begin{equation}
    \begin{split}
    \Phi_{E_2,d}(\vec R,\vec r) = \, &
        \delta_{\vec R,\vec \rho_A\pm\vec\delta_2}
        \delta_{\vec r,\pm\vec a_2}
         -
        \delta_{\vec R,\vec \rho_B\pm\vec\delta_1}
        \delta_{\vec r,\pm\vec a_1}.
    \end{split}
\label{eq:pi_E2_d}
\end{equation}
is not symmetry equivalent to the $\Phi_{E_2,3}$ and $\Phi_{E_2,\Bar{3}}$ states since the CMC coordinate always lies in the hexagon center of the kagome lattice, i.e. an empty site. Due to this we call it the $E_2,d$ where the $d$ stands for \textit{diagonal}.
Another difference to the $\Phi_{E_2,3}/\Phi_{E_2,\Bar3}$ states, is that the pairing sites have two intermediate sites as opposed to one, which decreases the relative contribution of the $\Phi_{E_2,d}$ state.
The associated states with opposite transformation behavior under mirroring along the $y$-axis complete the basis set for the $E_2$. These will be denoted by $\Phi_{E_2,i'}$. For
the onsite state $\Phi_{E_2,0}$ we get
\begin{equation}
     \Phi_{E_2,0'}(\vec R, \vec r) = \delta_{\vec r, \vec 0}
     \Big(
        +2\delta_{\vec R, \vec \rho_C}
        -\delta_{\vec R, \vec \rho_B}
        -\delta_{\vec R, \vec \rho_A}
     \Big) \,\text{.}
\label{eq:pi_E2_0p}
\end{equation}
The corresponding orthogonal states to $\Phi_{E_2,1}$ and $\Phi_{E_2,2}$ read
\begin{align}
    \Phi_{E_2,1'}(\vec R,\vec r) &=
        2 \delta_{\vec r, \pm\vec \delta_0}
        -
        \delta_{\vec r,\pm\vec \delta_1}
        -
        \delta_{\vec r,\pm\vec \delta_2} \,\text{,}\\
        \Phi_{E_2,2'}(\vec R,\vec r) &=
        2 \delta_{\vec r, \pm(\vec \delta_1+\vec\delta_2)}
        -
        \delta_{\vec r,\pm(\vec \delta_0+\vec\delta_1)}
        -
        \delta_{\vec r,\pm(\vec\delta_0-\vec \delta_2)} \,\text{.}
\label{eq:pi_E2_1p}
\end{align}
%Again, both the nearest and the next nearest neighbor states are structurally very similar.
The $\Phi_{E_2,3'}$ state, which transforms even under mirroring along the $y$-axis, has the same properties as it's orthogonal counterpart in that it does not factorize into CMC and RC components
\begin{equation}
    \begin{split}
    \Phi_{E_2,3'}(\vec R, \vec r) =&
        +2\,
        \big(
        \delta_{\vec R,\vec \rho_A}
            \delta_{\vec r,\pm\vec a_1} +
            \delta_{\vec R,\vec \rho_B}
            \delta_{\vec r,\pm\vec a_2}
        \big)\\
        &-\phantom{2}\,
        \big(
            \delta_{\vec R,\vec \rho_C}
            \delta_{\vec r,\pm\vec a_2}+
            \delta_{\vec R,\vec \rho_A}
            \delta_{\vec r,\pm\vec a_0}
        \big)
        \\
        &-\phantom{2}\,
        \big(
        \delta_{\vec R, \vec \rho_B}
        \delta_{\vec r,\pm\vec a_0}+
        \delta_{\vec R,\vec \rho_C}
        \delta_{\vec r,\pm\vec a_1}
        \big).
    \end{split}
\label{eq:pi_E2_3p}
\end{equation}
The same holds for $\Phi_{E_2,\Bar3'}$
\begin{equation}
	\begin{split}
		\Phi_{E_2,\Bar3'}(\vec R, \vec r) =&
		+2\,
		\big(
		\delta_{\vec R,\vec \rho_A}
		\delta_{\vec r,\pm\vec a_1} -
		\delta_{\vec R,\vec \rho_B}
		\delta_{\vec r,\pm\vec a_2}
		\big)\\
		&-\phantom{2}\,
		\big(
		\delta_{\vec R,\vec \rho_C}
		\delta_{\vec r,\pm\vec a_2}-
		\delta_{\vec R,\vec \rho_A}
		\delta_{\vec r,\pm\vec a_0}
		\big)
		\\
		&-\phantom{2}\,
		\big(
		\delta_{\vec R, \vec \rho_B}
		\delta_{\vec r,\pm\vec a_0}-
		\delta_{\vec R,\vec \rho_C}
		\delta_{\vec r,\pm\vec a_1}
		\big),
	\end{split}
	\label{eq:pi_E2_3Barp}
\end{equation}
which transforms odd under a mirror operation along the $y$-axis.
Lastly we have the even counterpart to $\Phi_{E_2,d}$
\begin{equation}
    \begin{split}
    \Phi_{E_2,d'}(\vec R,\vec r) = \, &
        +2\delta_{\vec R,\vec \rho_C\pm\vec\delta_0}
        \delta_{\vec r,\pm\vec a_0}\\
        &-\phantom{2}
        \delta_{\vec R,\vec \rho_B\pm\vec\delta_1}
        \delta_{\vec r,\pm\vec a_1}\\
        &-\phantom{2}
        \delta_{\vec R,\vec \rho_A\pm\vec\delta_2}
        \delta_{\vec r,\pm\vec a_2}.
    \end{split}
\label{eq:pi_E2_dp}
\end{equation}
which also does not factorize into CMC and RC.
The real space representations of the states are shown in \autoref{fig:evenStates}, \autoref{fig:E22States}.

\section{Analysing instabilities in the extended zone scheme}

\subsection{Extended zone scheme}\label{sec:extendedBZ}

The conventional approach to a structure with sublattice degree of freedom is
to treat the modulations periodic with the Bravais vectors in reciprocal space
within the corresponding BZ, while the sublattice structure is captured in an
additional orbital degree of freedom. Despite this approach is favourable for
both analytical and numerical calculations, it introduces an imbalance between
intra- and inter-unit cell modulations.
To remedy this shortcoming, one can reinterpret the Kagome lattice as an $1/4$
depleted triangular lattice with half the lattice constant.
The reciprocal lattice vectors double, and therefore the BZ quadruple in size,
consequently encoding the full spatial information of the sublattice without the
introduction of additional (orbital) degrees of freedom.
We refer to this enlarged BZ as the \textit{extended} BZ (eBZ).
Wave vectors within this eBZ carry the whole spatial information of the basis
sites, as it is also probed in experimental setups.
In contrast to a Bravais lattice with point group $C_{6v}$, the eBZ takes into
account the full space group of the Kagome lattice, which inclides
\textit{e.g.} a glide reflection (translation of half a Bravais vector and
subsequent inversion at this line). Breaking of this symmetry will lead to
finite $Q$ signals in experiments as discussed below.

To establish direct contact with the experiment and also gain deeper insight
into the real space structure of the superconducting pairing wave function,
one has to transfer the sublattice information from the sublattice degrees to
the eBZ of the spatially fully resolved lattice.
Since both perspectives encode the same information about the system, they are
connected by a unitary transformation, that we will establish in the
following.

\subsection{Unfolding susceptibilities into the eBZ}

To exemplify the transformation to the extended BZ utilising a momentum space picture,
we consider the following particle-particle susceptibility
\begin{equation}
\label{eq:PP_period}
\begin{split}
    \Pi^{ff'}_{\{o_i\}}\big(\mathbf{Q}\big) = & \,
    \frac{1}{N} \sum_{\mathbf{k},\mathbf{k}'}
    \varphi_f(\mathbf{k})
    \varphi^*_{f'}(\mathbf{k}') \\
    & \Big\langle
        c^\cre_{\mathbf{k} + \mathbf{Q},o_2}
        c^\cre_{-\mathbf{k},o_3}
        c^\ann_{-\mathbf{k}',o_1}
        c^\ann_{\mathbf{k}' + \mathbf{Q},o_0}
    \Big\rangle
\end{split}
\end{equation}
as the outcome of our TUFRG analysis.
The secondary momentum dependence in $\mathbf k$ and $\mathbf k^\prime$ was expanded
in a series of suitable formfactors $\varphi_f(\mathbf k)$.
As outlined in a previous section, this quantity is most conveniently obtained for a triangular lattice
with tri-atomic basis in the natural gauge~\cite{RevModPhys.84.299,Beyer2022},
which encapsulates all information about the real space position of the sublattices
relative to the unit-cell origin in orbital-like indices $o\in 
\{\text{A},\text{B},\text{C}\}$.
To reconstruct the spacial information in the momentum domain and establish a link to 
experimental probes for the superconducting gap function, one has to apply the gauge transformation
\begin{equation}
    c^\ann_{\mathbf k,o} \rightarrow c^\ann_{\mathbf k,o} e^{i \mathbf k \bm\gamma_o}
    \ ,
\label{eqn:gauge}
\end{equation}
which respects the actual positions of the sublattices $\bm\gamma_o$ corresponding to the sublattice site $o$.
Since the crystal momentum in Eq.~\ref{eq:PP_period} is defined with respect to the triangular superlattice,
the phase winding introduced by the sublattice positions $\bm\gamma_i$ is not
periodic in the initial- but in the extended BZ.
However, the Kagome lattice does not feature the full periodicity in the triangular lattice 
vectors $\Bdelta_0 = \mathbf a_0 / 2$ and $\Bdelta_1 = \mathbf a_1 / 2$,
since $1 / 4$ of the lattice sites remain empty.
This reduced symmetry allows a finite support in the extended BZ only for harmonic functions
without occupation on the vacant sites.
This is equivalent to a FT of a triangular lattice of half the spacing with a
nontrivial structure factor as known from diffraction
experiments~\cite{Ashcroft1976}.
Applying the gauge transformation of Eq.~\ref{eqn:gauge} to the inverse of Eq.~\ref{eq:PP_period} and expanding
the result in a set of formfactors adapted to the extended BZ $\varphi_g(\mathbf k)$, one obtains
\begin{equation}
\begin{split}    
    \Pi^{gg'}\big(\mathbf{Q}\big) = &
    \sum_{f,f'} \sum_{\{o_i\}}
    e^{i\mathbf{Q}(\bm\gamma_{o_0}-\bm\gamma_{o_2})} 
    \Pi^{ff'}_{\{o_i\}}\big(\mathbf{Q}\big)
    T^{gg'ff'}_{\{o_i\}}
\end{split}
\label{eq:eBZ_trafo}
\end{equation}
with the momentum independent transformation matrix
\begin{equation}
\begin{split}
    T^{gg'ff'}_{\{o_i\}} =
    \sum_{\mathbf{k}, \mathbf{k}'} & \,
    e^{i\mathbf{k}(\bm\gamma_{o_3}-\bm\gamma_{o_2})}
    e^{i\mathbf{k}'(\bm\gamma_{o_0}-\bm\gamma_{o_1})} \\
    & \phi_g(\mathbf{k})
    \phi^*_{g'}(\mathbf{k}')
    \varphi^*_f(\mathbf{k})
    \varphi_{f'}(\mathbf{k}') \ .
\end{split}
\end{equation}
This susceptibility presents the full Fourier transformed SC pairing
susceptibility, that can be directly probed in Josephson STM measurements with
a superconducting tip~\cite{Pan1998, PhysRevB.64.212506,  Proslier2006, PhysRevB.80.144506}.
It encodes all information about the spatial ordering in the momentum distribution
and emerges from the convolution of the orbital dependent susceptibility
$\Pi^{ff'}_{\{o_i\}}\big(\mathbf{Q}\big)$ with the atomic structure factor
$e^{i\mathbf{Q}(\bm\gamma_{o_0}-\bm\gamma_{o_2})}$.
An analogous procedure can be applied to particle hole susceptibilities to
characterise magnetic and charge ordering.

Apparently, both description are equivalent, since they only involve a gauge transformation.
This can be easily understood by choosing plane wave formfactors
$\varphi_f(\mathbf k) = e^{i \mathbf k \latVec_f}$ with $\latVec_f$ a real space lattice vector.
Upon reducing the unit cell to a single site, the number of formfactors with finite support
$n_g$ is equivalent to the number of sublattice positions in the original unit cell
$n_o$ times the number of previously employed formfactors $n_f$.
We therefore see, that the intra-unit cell degrees of freedom are faithfully transferred to the corresponding momentum structure without any loss of information.
Assuming
\begin{equation}
    \varphi_f(\mathbf{k}) = e^{-i\mathbf{k}\mathbf{R}_f}
    \qquad \phi_g(\mathbf{k}) = e^{-i\mathbf{k}\mathbf{r}_g}
\end{equation}
we find
\begin{equation}
    \begin{split}
        t_{o_2o_3}^{fg} &= 
        \delta(\Bdelta_3-\Bdelta_2-\mathbf{r}_g+\mathbf{R}_f) \\
        t_{o_0o_1}^{f'g'} &= 
        \delta(\Bdelta_0-\Bdelta_1-\mathbf{R}_{f'}+\mathbf{r}_{g'})
        \,\text{,}
    \end{split}
\end{equation}
and clearly 
\begin{equation}
    T^{gg'ff'}_{\{o_i\}} = t_{o_2o_3}^{fg}t_{o_0o_1}^{f'g'} \ ,
\end{equation}
\textit{i.e.} $T$ confines contributing formfactors to the actually occupied
sites in the real space lattice structure.

\begin{figure}
    \centering
\includegraphics[width=0.5\textwidth]{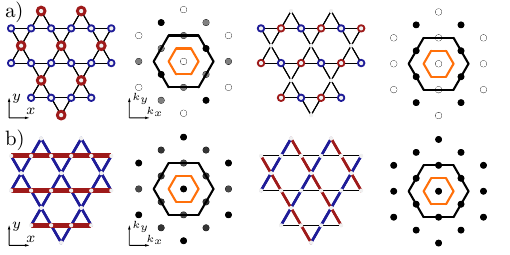}
    \caption{Intra unit cell particle particle instabilities transforming under the $E_2$ irreducible representation.
    Considering only pairing of electrons up to the nearest neighboring sites, the number of pairing functions within the $E_2$ irreducible representation is limited to 4: a) shows the real space representations of the onsite pairing states leading to a SMS order. b) represents the nearest neighbor bond order state correspond to a $d$-wave superconductor.
    Red (blue) sites (bonds) encode a positive (negative) value of the real valued gap function.
    While both orders reside at $\mathbf Q = 0$ in the reduced BZ (orange), they can be distinguished in the extended BZ (black).
    }
    \label{fig:E2_states}
\end{figure}

\subsection{SMS order and $d$SC in the eBZ}

As discussed in the main text, there are two ways to achieve a pairing
structure in the Kagome lattice, which follows $d$-wave symmetries.
As for any lattice, a nearest neighbor pairing with suitable amplitudes can be constructed.
However, because of the three basis points within one Kagome unit-cell arranged in a $3/4$ depleted 
triangular lattice,
there exist an additional possibility of having a modulation in the on-site pairing
amplitude without disrupting the translational symmetry of the underlying Bravais lattice.
In the conventional reduced zone scheme, both states therefore will feature an
ordering vector at the $\Gamma$ point.

The first choice, as expected from a superconducting state, is completely uniform
along all direction of the nearest neighbor bonds, while the latter clearly is not.
This will result in an ordering vector away from $\Gamma$ in the extended zone scheme.
To illustrate this, compare the two states with $(-1, -1, 2)$ modulation as depicted
in \autoref{fig:E2_states}. The real-space pairing is given by
Eq.~\eqref{eq:pi_E2_0p} for the modulated on-site pairing and
Eq.~\eqref{eq:pi_E2_1p} for the translationally invariant nearest-neighbor pairing.
Let us first consider the momentum structure of these pairings in the reduced scheme, given by
\begin{equation}
\begin{split}
    \Phi^{o_0o_1}(\mathbf{Q},\mathbf{k}) = \sum\limits_{\mathbf{r}_0, \mathbf{r}_1}
    \Phi(\mathbf{r}_0, \mathbf{r}_1)
    e^{-i\mathbf{Q}(\mathbf{r}_0 - \Bdelta_{o_0} + \mathbf{r}_1- \Bdelta_{o_1})/2}\\
    e^{-i\mathbf{k}(\mathbf{r}_1- \Bdelta_{o_1} - \mathbf{r}_0 + \Bdelta_{o_1})}.
\end{split}
\end{equation}
Substituting \eqref{eq:pi_E2_0p}, we find
\begin{equation}
    \begin{split}
        \Phi_{E_2,0^\prime}^{o_0o_1}(\mathbf{Q},\mathbf{k})
    = \sum\limits_{\mathbf{R}_0,\mathbf{R}_1} &\delta_{\mathbf{R}_0, \mathbf{R}_1} 
        \delta_{o_0,o_1} e^{-i\mathbf{Q}(\mathbf{R}_0 + \mathbf{R}_1)/2} \\
    &(-\delta_{o_0,0}-\delta_{o_0,1}+2\delta_{o_0,2}),
    \end{split}
\end{equation}
where $\mathbf{R}_i$ are the Bravais vectors of the unit cell $\mathbf{r}_i$ resides in.
Using the completeness of lattice Fourier transforms, this simplifies to
\begin{equation}
    \begin{split}
    \Phi_{E_2,0^\prime}^{o_0o_1}(\mathbf{Q},\mathbf{k}) =  &\delta_{o_0,o_1}(-\delta_{o_0,0}-\delta_{o_0,1}+2\delta_{o_0,2}) \delta_{\mathbf{Q},\mathbf{G}}\\
    = &\begin{pmatrix}
        -1 & 0 & 0\\
        0 & -1 & 0\\
        0 & 0 & 2
    \end{pmatrix} \delta_{\mathbf{Q},\mathbf{G}}
    \end{split}
\end{equation}
with a reciprocal lattice vector $\mathbf{G}$ of the Kagome lattice.
Within the BZ, therefore, this gap function has support only at $\mathbf{Q} = \Gamma$, with uniform distribution in $\mathbf{k}$.
Repeating the same procedure for $\Delta_{\mathrm{dSC}}$ we find
\begin{equation}
    \begin{split}
    &\Phi_{E_2,1^\prime}^{o_0o_1}(\mathbf{Q},\mathbf{k}) =
    \delta_{\mathbf{Q},\mathbf{G}} \\
    &\begin{pmatrix}
        0 & 2(1+e^{-i \mathbf{k} \mathbf{a_1}}) & -(1+e^{-i \mathbf{k} \mathbf{a_2}})\\
        2(1+e^{i \mathbf{k} \mathbf{a_1}}) & 0 & -(1+e^{i \mathbf{k} (\mathbf{a_1}-\mathbf{a_2})})\\
        -(1+e^{i \mathbf{k} \mathbf{a_2}}) & -(1+e^{-i \mathbf{k} (\mathbf{a_1}-\mathbf{a_2})}) & 0
    \end{pmatrix} , 
\end{split}
\end{equation}
again with support only at $\mathbf{Q} = \Gamma$ within the BZ but now with a
structure in $k$.

Shifting to the eBZ picture, we have to take the real--space positions of the basis points into consideration. The gap function in this representation in reciprocal space is given by
\begin{equation}
    \begin{split}
            \Phi(\mathbf{Q},\mathbf{k}) = &\sum\limits_{\mathbf{r}_0,
            \mathbf{r}_1} \Phi(\mathbf{r}_0, \mathbf{r}_1)
            e^{-i\mathbf{Q}(\mathbf{r}_0 + \mathbf{r}_1)/2}
            e^{-i\mathbf{k}(\mathbf{r}_1 - \mathbf{r}_0)}\\
            = &\sum\limits_{o_0,o_1} \Phi^{o_0o_1}(\mathbf{Q},\mathbf{k})  e^{-i\mathbf{Q}( \Bdelta_{o_0} +  \Bdelta_{o_1})/2}  e^{-i\mathbf{k}(\Bdelta_{o_0} - \Bdelta_{o_1})}
    \end{split}
\end{equation}
according to Eq.~\ref{eq:eBZ_trafo}.
Consequently, the eBZ expression for the on-site pairing is given by
\begin{equation}
    \Phi_{E_2,0^\prime}(\mathbf{Q},\mathbf{k}) = (-1-e^{-i\mathbf{Q}\Bdelta_0}+2e^{-i\mathbf{Q}\Bdelta_1})\delta_{\mathbf{Q},\mathbf{G}}.
\end{equation}
Because, in the Kagome lattice the sublattice vectors $\Bdelta_i = \mathbf a_i/2$
are half of the Bravais vectors, the phase factors will result in multiples of
$\pi$ for the reciprocal lattice points, such that there will be support on
the M points of the eBZ, but not at $\Gamma$.

For the uniform state, in contrast, we find
\begin{equation}
\begin{split}
    \Phi_{E_2,1^\prime}(\mathbf{Q},\mathbf{k})
    = &(2\cos(\mathbf{k}\Bdelta_0)e^{-i \mathbf{Q} \Bdelta_0/2}\\
    - &\cos(\mathbf{k}\Bdelta_1)e^{-i \mathbf{Q} \Bdelta_1/2}\\
    - &\cos(\mathbf{k}(\Bdelta_0-\Bdelta_1))e^{-i \mathbf{Q} (\Bdelta_0-\Bdelta_1)/2})
    \delta_{\mathbf{Q},\mathbf{G}} \ .
\end{split}
\end{equation}
While this state also features some dependence on the relative momentum
$\mathbf k$ of the Cooper pair constituents, for fixed $\mathbf k$ the phases
associated with the Cooper pair momentum $\mathbf Q$ are not able to
destructively interfere in order to suppress a peak in the eBZ.
This results in the distinct $\mathbf Q$ map of the SMS and $d$SC phase
depicted in \autoref{fig:E2_states}.
Variations in gap size as well as the phase difference of the superconducting order parameter could be probed at atomic
scale within experiments using phase sensitive superconducting STM techniques \cite{Karan_2022}.

\section{Candidate state analysis of FRG results}
In order to trace the competing contributions within the FRG flow, i.e. as a function of energy cutoff $\Lambda$, we determine the expectation value
\begin{equation}
\Xi \,=\, \braket{\Phi_{\nu,\theta} |\, V_{\nu} \,| \Phi_{\nu,\theta} }
\end{equation}
of the effective FRG vertex $ V_{\nu} $ at each flow step. \mbox{$\nu \in\,
$(pp, cha, mag)} denotes the physical instability channels reconstructed from
the TUFRG decomposition in Eq.~\eqref{eq:PP_vertex}, \eqref{eq:Charge_vertex},
\eqref{eq:Spin_vertex}.
$\theta$ defines the candidate state, for which we take the final FRG state of the respective phases.

To calculate the overlaps of the FRG state with the candidate states we define
\begin{equation}
\label{eq:projection}
\gamma_{E_2,i}=|\braket{\Psi_\text{FRG}|\Phi_{E_2,i}}|^2+|\braket{\Psi_\text{FRG}|\Phi_{E_2,i'}}|^2.
\end{equation}
As shown in \autoref{tab:expVals}, the FRG state is predominantly composed of $\Phi_{E_2,0}$ and $\Phi_{E_2,3}$.
When increasing the nearest-neighbor repulsion $V$, the ratio $\Phi_{E_2,3} / \Phi_{E_2,0}$ declines as longer-ranged fluctuations are suppressed.
To highlight this fact we define the ratio
\begin{equation}
\label{eq:ratio}
\alpha \,=\, \dfrac{1}{2} + \dfrac{1}{2} \, \dfrac{\gamma_{E_2,0} - \gamma_{E_2,3}}
    {\gamma_{E_2,0} + \gamma_{E_2,3}}
\end{equation}
resulting in $\alpha=0$ ($\alpha=1$) for a pure $\Phi_{E_2,3}$ ($\Phi_{E_2,0}$) contribution.

\begin{table}[H]
\begin{center}
\begin{tabular}{ |c||c|c|c|c|c|c|c|c| }
 \hline
 $V/t$ & $\gamma_{E_2,0}$ & $\gamma_{E_2,1}$ & $\gamma_{E_2,2}$ & $\gamma_{E_2,3}$ & $\gamma_{E_2,d}$ & $\gamma_{E_2,\Bar3}$& $\gamma_{A_1}$ & $\Sigma$ \\ [0.5ex]
 \hline
 \hline
0.50 & 50.86 & 0.00 & 0.55 & 43.76 & 0.11 & 0.01 & 0.0 & 95.31\\
0.52 & 53.75 & 0.00 & 0.60 & 41.47 & 0.04 & 0.01 & 0.0 & 95.87\\
0.54 & 56.13 & 0.00 & 0.63 & 39.44 & 0.01 & 0.01 & 0.0 & 96.23\\
0.56 & 58.18 & 0.00 & 0.66 & 37.63 & 0.00 & 0.01 & 0.0 & 96.49\\
0.58 & 59.95 & 0.00 & 0.69 & 36.02 & 0.00 & 0.01 & 0.0 & 96.68\\
0.60 & 61.53 & 0.00 & 0.72 & 34.57 & 0.01 & 0.01 & 0.0 & 96.84\\
0.62 & 62.92 & 0.00 & 0.74 & 33.25 & 0.03 & 0.01 & 0.0 & 96.97\\
0.64 & 64.18 & 0.00 & 0.77 & 32.05 & 0.06 & 0.02 & 0.0 & 97.07\\
0.66 & 65.29 & 0.01 & 0.80 & 30.96 & 0.09 & 0.02 & 0.0 & 97.16\\
0.68 & 66.30 & 0.01 & 0.82 & 29.95 & 0.13 & 0.02 & 0.0 & 97.22\\
\hline
\hline
0.70 & 0.0 & 0.0 & 0.0 & 0.0 & 0.0 & 0.0 & 80.21 & 80.21\\
0.72 & 0.0 & 0.0 & 0.0 & 0.0 & 0.0 & 0.0 & 80.82 & 80.82\\
\hline
\end{tabular}
\end{center}
\caption{Overlap of the final FRG state (pp channel) with candidate states at $U=0.14$ in percent. Values are calculated according to \autoref{eq:projection}.  }
\label{tab:expVals}
\end{table}

\section{SMS Mean field results}

\begin{figure}
    \centering
\includegraphics[width=\columnwidth]{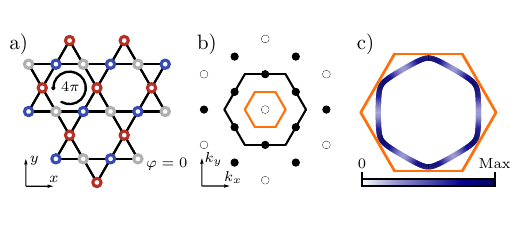}
    \caption{Kagome sublattice modulated SC (SMS) order at U/t=1.0, V/t=1.1.
    The SMS state is dominated by on-site correlations that display a phase modulation in the unit cell as shown in a). 
    The phase of the local order parameter $\Phi(\mathbf{R}_S,\mathbf{0})$ is indicated by the color of the dots where blue (red) represents $\varphi=2\pi/3$ ($\varphi=-2\pi/3$) and white depicts $\varphi=0$. The absolute value is identical for all points.
    The representation in the extended zone scheme in b) coincides with the expected $\mathbf{k}$ space patterns accessible in STM measurements.
    c) shows the absolute value of the SMS gap on the FS.
		}
    \label{fig:measurement}
\end{figure}

In the case of the onsite $E_2$ states, dominating the SMS domain, the
free energy is minimized within the MF approximation by the chiral
superposition of $\Phi_{E_2, 0^{(\prime)}}$ depicted in
Fig.~\ref{fig:measurement}a).
This leads to a phase winding of $4 \pi$ within each hexagon with equal weights
on the sublattices.
The phases are transferred to the peak structure within the eBZ in
Fig.~\ref{fig:measurement}b).
While the absolute values restore the hexagonal symmetry, the phase modulation
along each kagome chain still results in a finite Cooper pair momentum with
respect to the single-site unit cell.
This linear combination of degenerate $E_2$ states fully gaps the FS with
larger values at the $M$ points, where density of states accumulates.
The gap structure is almost identical to that of a chiral $d$-wave SC, which
omitted a correct identification of the SMS state in previous
works~\cite{PhysRevB.87.115135}. As outlined before, the modulation of the phase
can be probed within STMs with superconducting tips \cite{Karan_2022}, enabling the distinction between ordinary 
superconductors and the proposed new type of SMS.

\bibliography{pdw_SM}